\documentclass[final,5p,times,twocolumn]{elsarticle}

\usepackage{amssymb}

\usepackage{color}
\usepackage{amsmath}
\usepackage{multirow}
\usepackage{booktabs}
\usepackage{bm}
\usepackage{braket}
\usepackage{amsmath}
\usepackage{graphicx}
\biboptions{sort&compress}

\usepackage[pdfstartview=FitH, colorlinks,
 linkcolor={red!60!black!},
 anchorcolor={green!80!black!},
 urlcolor={blue!80!black!},
 citecolor={blue!50!black!}]{hyperref}
\usepackage[table]{xcolor}

\journal{Physics Letters B}

\begin{document}


\begin{frontmatter}



\title{One-neutron halo structure of $^{29}$Ne}


\author[ad1,ad2]{J. G. Li\corref{correspondence}}
\author[ad1,ad2]{N. Michel\corref{correspondence}}
\author[ad1,ad2]{H. H. Li}
\author[ad1,ad2]{W. Zuo\corref{correspondence}}

\address[ad1]{CAS Key Laboratory of High Precision Nuclear Spectroscopy, Institute of Modern Physics,
Chinese Academy of Sciences, Lanzhou 730000, China}
\address[ad2]{School of Nuclear Science and Technology, University of Chinese Academy of Sciences, Beijing 100049, China}

\cortext[correspondence]{Corresponding author.
e-mail address: jianguo\_li@impcas.ac.cn (J. G. Li), nicolas.michel@impcas.ac.cn (N. Michel), zuowei@impcas.ac.cn (W. Zuo)}

\begin{abstract}
  We have applied the Gamow shell model to calculate nuclear observables of $^{26-31}$Ne isotopes pertaining to one-neutron halo structure, these nuclei being situated close to neutron drip-line.
  As both many-body correlations and continuum coupling are taken into account in that approach, halo structure can be analyzed properly.
  Our calculations provide good descriptions of $^{26-31}$Ne, where asymptotic behavior is crucial for that matter.
  One-body density, neutron root-mean-square radii of $^{26-31}$Ne, and one-neutron overlap functions of $^{29,31}$Ne have been calculated as well.
  Our results support the presence of a one-neutron \textit{p}-wave halo in $^{31}$Ne, already pointed out experimentally.
  A similar situation also occurs in the ground state of $^{29}$Ne, which is mainly a \textit{p}-wave valence neutron coupled to the inner $^{28}$Ne \textit{core}.
  The $3/2^+$ excited state of $^{29}$Ne, which is dominated by a \textit{d}-wave valence neutron, has also been considered.
  A larger radius and more extended wave function occur for the ground state of $^{29}$Ne when compared to its $3/2^+$ first excited state.
  The present results suggest that $^{29}$Ne is a good candidate for one-neutron \textit{p}-wave halo in the medium-mass region.

\end{abstract}

\begin{keyword}
One-neutron halo \sep Continuum coupling \sep Gamow shell model \sep Neutron rms radius \sep Overlap function \sep Spectroscopic factor
\end{keyword}

\end{frontmatter}



\textit {Introduction.} --
 Advanced new radioactive ion beam facilities dedicated to the study of drip-line nuclei have provided access to proton-rich and neutron-rich exotic nuclei which could not be synthesized with accelerators of older generation  \cite{Blumenfeld_2013}.
 The proton drip-line has been experimentally delineated up to \textit{Z} = 93 \cite{PhysRevLett.122.192503}, but the neutron-drip-line nuclei could be identified only up to $^{39}$Na and $^{40}$Mg \cite{Baumann20071022,PhysRevLett.123.212501}.
 New-generation radioactive isotope beam facilities, such as FRIB \cite{frib} and HIAF \cite{HIAF}, are being constructed in order to explore and discover the new physics  at nuclear drip-lines.
 Contrary to well-bound nuclei, which are closed quantum systems, drip-line nuclei are open quantum systems, as they can be either weakly bound or unbound with respect to particle-emission threshold \cite{0954-3899-36-1-013101}.
 Interesting phenomena appear near nuclear drip-lines, such as halo structure, especially in neutron-rich nuclei  \cite{TANIHATA2013215}.
 In fact, drip-line nuclei offer unique laboratories to better understand many-body correlations and nuclear forces subject to large neutron-to-proton ratios.
 Nuclear halo is a threshold effect arising from the least bound one or two valence nucleons, which decouple from an inner \textit{core} containing all the other nucleons \cite{TANIHATA2013215}.
 At nuclear drip-lines, halos are not only found in the lightest systems, but also occur in nuclei as heavy as $^{37}$Mg \cite{TANIHATA2013215,PhysRevC.90.061305}.
 Halo nuclei are characterized by a nucleon density extending to a large distance, where the tail of nucleon distributions slowly decreases at large distance.
 This corresponds to a significant probability to find one or two nucleons at large radius \cite{TANIHATA2013215,PhysRevC.101.031301}.

 Several halo nuclei have been identified in light mass region, such as the two-neutron halo nuclei $^{6,8}$He \cite{PhysRevLett.93.142501,PhysRevLett.99.252501}, $^{11}$Li \cite{PhysRevLett.55.2676}, $^{17,19}$B  \cite{PhysRevLett.126.082501,PhysRevLett.124.212503} and $^{22}$C \cite{PhysRevLett.104.062701}
 as well as the one-neutron halo nuclei $^{11}$Be \cite{PhysRevLett.74.30} and  $^{15,19}$C \cite{PhysRevC.69.034613,PhysRevLett.117.102501}.
 Only a few halo nuclei are known in the medium mass region, such as the one-neutron halo present in $^{31}$Ne \cite{PhysRevLett.109.202503,PhysRevLett.112.142501} and $^{37}$Mg \cite{PhysRevLett.112.242501},
 and the two-neutron halo in $^{29}$F \cite{PhysRevLett.124.222504}. The existence of halo structure implies that significant modifications in the shell structure of these nuclei are generated.
 In particular, they involve mixing in their ground states induced by intruder states, such as \textit{s} or \textit{p} partial waves \cite{TANIHATA2013215}.
 Intruder state mixing can lead to considerable deformations in these states. Coupling to higher partial waves via configuration mixing can also occur.
 Hence, continuum coupling and many-body correlations must be treated properly for a proper description of halo nuclei \cite{TANIHATA2013215}.
 In Gamow shell model (GSM) \cite{PhysRevLett.89.042502,PhysRevLett.89.042501,0954-3899-36-1-013101,Michel_GSM_book,physics3040062,PhysRevC.103.034305},
 both continuum coupling and many-body correlations are included, so that it is an appropriate tool to precisely study the exotic properties of drip-line nuclei, in particular halo structures \cite{PhysRevC.84.051304}.
 In fact, GSM \cite{PhysRevLett.89.042502,PhysRevLett.89.042501,0954-3899-36-1-013101} has become a powerful and predictive tool to determine the nuclear structure and reaction observables of drip-line nuclei \cite{SUN2017227,PhysRevC.94.054302}.

 Neon isotopes have been synthesized up to the neutron drip-line, as it is reached via the drip-line nucleus $^{34}$Ne \cite{PhysRevLett.109.202503,Wang_2021}.
 Its rich structure has also been established experimentally \cite{PhysRevC.93.014613,PhysRevLett.112.142501,LIU201758,PhysRevC.99.011302}.
 The neon chain forms an interesting ground for theoretical studies, as they can provide information to understand the proton-neutron and neutron-neutron interactions, as well as many-body correlations at drip-lines.
 Moreover, $^{31}$Ne is a one-neutron halo \cite{PhysRevLett.112.142501,PhysRevC.96.064603}, as the valence neutron of $^{31}$Ne is found to occupy \textit{p}-waves, instead of  \textit{f}-waves,
 the latter situation being expected from standard shell model \cite{sym13112167}.
 As it has a small one-neutron separation energy, of about 200 keV, a one-neutron halo can form due to  the coupling of a $p$-wave valence neutron coupled with the deformed $^{30}$Ne inner \textit{core}.
 Added to that, the $3/2^-$ ground state of $^{29}$Ne exhibits large \textit{p}-wave intruder configuration components \cite{PhysRevC.93.014613,LIU201758}
 in conjunction to the slight deformation found in the ground state of $^{28}$Ne \cite{PhysRevC.93.014613,PhysRevC.99.011302}.  Conversely, the first $3/2^+$ excited state of $^{29}$Ne is dominated by localized configurations.
The one-neutron separation energy ($S_n$) of $^{29}$Ne is about 900 keV \cite{PhysRevC.93.014613}, which is larger than the one-neutron separation energy $S_n$ of $^{31}$Ne, which is 200 keV \cite{PhysRevLett.112.142501,Wang_2021}.
However, $S_n$ in $^{29}$Ne is comparable to the one-neutron \textit{s}-wave halo nuclei $^{15,19}$C \cite{Wang_2021}, which are about 1200 and 700 keV, respectively.
Unfortunately, contrary to $^{31}$Ne, there is scarce experimental and theoretical information available on the ground state of $^{29}$Ne, which is likely a one-neutron halo.
The similar structures of $^{29}$Ne and $^{31}$Ne can surely give rise to interesting phenomena and shed light on the suspected one-neutron halo properties of $^{29}$Ne.

In GSM,  many-body correlations  are included  via configuration mixing and continuum coupling is present at basis level \cite{0954-3899-36-1-013101}.
Therefore, GSM is suitable for the description of exotic properties in drip-line nuclei. It is the object of the present paper to study neon isotopes situated close to neutron drip-line with GSM, in order to investigate the halo properties of $^{31}$Ne and $^{29}$Ne.
The present paper is written as follows. 
The basic framework of GSM is firstly introduced briefly. Then, we present the model space and many-body Hamiltonian of GSM used in the present work. 
In order to ponder about the halo structures of $^{31}$Ne and $^{29}$Ne, we will show the one-body densities, neutron root-mean-square (rms) radii, and single-particle (s.p.) overlap functions of neon isotopes obtained from GSM calculations.  
Conclusion will be made afterwards.

\textit{Model.} --
The fundamental theoretical construction entering GSM is the one-body Berggren basis. The Berggren basis was initially proposed by T.~Berggren in Ref.\cite{BERGGREN1968265}.
It consists of bound, resonance and scattering s.p.~states, generated by a finite-range potential, such as Woods-Saxon (WS) potential. The completeness of the basis \cite{BERGGREN1968265} for a given partial wave of quantum number $\ell j$ reads :
\begin{equation}
\sum_n \ket{n \ell j} \bra{n \ell j} + \int_{L_+} \ket{k \ell j} \bra{k \ell j}~dk = 1 \label{Berggren}
\end{equation}
where $\ket{n \ell j}$ runs over resonant (bound or resonance) states, and $L_+$ is a complex contour in the complex-momentum space encompassing resonance states.
The $L_+$ integral contains the continuous part of the Berggren basis, with $\ket{k \ell j}$ running over the scattering states belonging to the $L_+$ contour.
In numerical GSM calculations, the $L_+$ integral is efficiently discretized with the Gauss-Legendre quadrature, as convergence is obtained with 20-30 points \cite{0954-3899-36-1-013101,Michel_GSM_book,physics3040062}. 
From the Berggren basis of Eq.~(\ref{Berggren}), a many-body basis of Slater determinants can be generated,
so that internucleon correlations are described via configuration mixing in a shell-model picture \cite{PhysRevLett.89.042502,PhysRevLett.89.042501,0954-3899-36-1-013101,Michel_GSM_book,physics3040062,MICHEL2020106978,Michel_GSM_book}. 
Hence, GSM is a suitable approach to study many-body weakly bound and unbound states.

GSM is usually performed in the picture of a core plus valence particles.
In order to properly account for center-of-mass degrees of freedom, one uses the cluster-orbital shell model framework (COSM) \cite{Suzuki1988}.
Nucleon COSM coordinates are defined with respect to the center-of-mass of the core, so that one works in a translationally invariant frame, thus where center-of-mass excitations cannot develop.
The GSM Hamiltonian using COSM reads :
\begin{eqnarray}\label{COSM_H}
\hat{H} = \sum_{i=1}^{N_{\text{val}}}  \left (\frac{\hat{\mathbf{p}}_i^2}{2\mu_i} + \hat{U}^{(c)}_i \right)+ \sum_{i<j}^{N_{\text{val}}} \left ( \hat{V}_{ij} + \frac{\hat{\mathbf{p}}_i \cdot \hat{\mathbf{p}}_j}{M_c} \right),
\end{eqnarray}
where $N_{\text{val}}$ is the number of valence nucleons, $\mu_i$ is the reduced mass of the $i$-th nucleon with respect to the core, and $M_c$ is the mass of the core. 
The one-body potential $\hat{U}_i$ is represented by a WS potential mimicking the inert core.
$V_{ij}$ is the two-body residual interaction, which is modeled for our present work by a pionless effective field theory (EFT) interaction \cite{CONTESSI2017839,RevModPhys.85.197,RevModPhys.92.025004}.
The part proportional to $\hat{\textbf{\emph p}}_i \cdot \hat{\textbf{\emph p}}_j$ is an additional two-body kinetic term, which takes into account the recoil of the active nucleons relative to the chosen core in the COSM framework.
Concerning the used pionless EFT interaction \cite{CONTESSI2017839,RevModPhys.85.197,RevModPhys.92.025004}, only two-body contact terms up to next-to-next leading-order are considered in the present calculations.
EFT parameters are optimized to reproduce the energies of the low-lying states of selected nuclei.
The harmonic oscillator (HO) basis used for the representation of the EFT interaction is limited to a few shells.
Thus, the high-momentum components of the initial EFT interaction are automatically suppressed.
This regularization approach has been recently utilized in Refs.~\cite{PhysRevC.93.044332,PhysRevC.86.031301,PhysRevC.98.054301,PhysRevC.98.044301}, and in particular in GSM with the calculation of neutron-rich oxygen isotopes \cite{PhysRevC.103.034305}.

\begin{table}[htb]
\centering
\caption{\label{Table.InterParamEFT} Optimized parameters of the EFT interaction at leading order (LO) and next-to-leading order (NLO). They are given in natural units.
The $C_S^{0,1},C_T^{0,1},C_{1 \dots 7}$ notations are taken from Refs.~\cite{CONTESSI2017839,RevModPhys.85.197,RevModPhys.92.025004} . The parameters at leading order ($C_S^{0,1},C_T^{0,1}$) explicitly depend on the isospin $T=0,1$ of the two nucleons.}
\setlength{\tabcolsep}{0.4mm}{
\begin{tabular}{ccccccccc} \hline \hline
LO constant &$C_S^{0}$   & $C_S^{1}$    & $C_T^{0}$  & $C_T^{1}$   \\  \hline 
LO value&  $-$0.071 &  $-$0.068  &  0 &  0 \\ 
NLO constant &$C_1$   & $C_2$    & $C_3$  & $C_4$  & $C_5$   & $C_6$  & $C_7$ \\ \hline 
NLO value&  0.16  &  $-$0.78  &  $-$0.05 &  0.45 & $-$1.74 & $-$0.29 & $-$0.02 \\ \hline \hline
\end{tabular}}
\end{table}

The closed-shell nucleus $^{24}$O is used as the inert core in this work.
As valence protons are well bound in neutron-rich neon isotopes, it is sufficient to consider the harmonic oscillator states 0$d_{5/2}$ and 1$s_{1/2}$ to generate the proton part of the model space.
For valence neutrons, the $d_{3/2}$, $f_{7/2}$, $p_{3/2}$ and $p_{1/2}$ partial waves are represented by the Berggren basis, as they are most crucial to generate a proper coupling to the continuum.
The parameters of the one-body WS potential and two-body EFT force in the GSM Hamiltonian are optimized to reproduce the low-lying states of $^{25,26}$O, $^{25,26}$F and $^{26-29}$Ne. More precisely, all the ground states and first excited states of these isotopes were fitted, except for the first excited state of $^{25}$O. The third and fourth excited states of $^{26}$F have also been added to the fit. The fixed parameters of the WS core potential are the diffuseness $d = 0.65$ fm, the radius $R_0 = 3.663 $ fm and the spin-orbit coupling $V_{\ell s} = 7.5$ MeV.
The central depth $V_0$ differs according to partial waves : it is equal to 65.81 and 62.21 MeV for proton $s_{1/2}$ and $d_{5/2}$ partial waves,
whereas, for the neutron part, it is 40 MeV for $\ell=0,2$, 47 MeV for $\ell=1$ and 38 MeV for $\ell =3$ partial waves.
The resulting fit of the parameters of the pionless EFT interaction is shown in Tab.~\ref{Table.InterParamEFT}.
The EFT interaction is separated into two parts: the leading-order part ($C_S^{0,1}$ and $C_T^{0,1}$) and the next-to-leading order ($C_{1...7}$) part
(see Refs.\cite{CONTESSI2017839,RevModPhys.85.197,RevModPhys.92.025004} for notation and definition of associated operators).
The $C_S$ and $C_T$ constants depend on the isospin $T=0,1$ of the two nucleons.
As $C_S$ and $C_T$ reduce to a single constant in our framework, we only fitted $C_S$ and arbitrarily set $C_T = 0$.
The obtained Hamiltonian provides the neutron separation energy of $^{29}$Ne to about 840 keV, which is consistent with its experimental value of 900 keV \cite{PhysRevC.93.014613}.
We also performed GSM calculations for $^{30,31}$Ne using the optimized Hamiltonian. $S_n$ in $^{31}$Ne is about 350 keV, which is also close to its experiment value of 200 keV.
Consequently, as we obtained a good agreement of the one-neutron separation energies $S_n$ in $^{29,31}$Ne with experimental data, we can proceed to the quantitative investigation of the halo properties of $^{29,31}$Ne.

\textit{Results.} --
The one-neutron \textit{p}-wave halo nucleus $^{31}$Ne has been experimentally studied using 1$n$-removal reactions \cite{PhysRevLett.112.142501}.
It was concluded that the one-neutron halo $^{31}$Ne isotope is deformation-driven, of $p$-wave character, and exhibits a small $S_n$ of  $150^{+160}_{-100}$ keV \cite{PhysRevLett.112.142501}.
Recent experiments \cite{PhysRevC.93.014613,LIU201758} showed that the ground state of $^{29}$Ne is dominated by $p$-wave intruder configurations with $J^\pi = 3/2^-$, where $S_n$ is about 900 keV.
Whether $^{29}$Ne exhibits one-neutron halo properties remains, however, to be determined.
Moreover, the inner \textit{core} $^{28}$Ne in $^{29}$Ne is not so much  deformed as the inner core $^{30}$Ne of the one-neutron halo nucleus $^{31}$Ne \cite{PhysRevC.93.014613,PhysRevC.99.011302}.
In fact, due to their subtle interplay of continuum coupling and deformation, the $^{29,31}$Ne isotopes offer unique prototype systems to study the mechanism of shell evolution and halo formation in the medium-mass region.

\begin{figure}[!htb]
\includegraphics[width=1.0\columnwidth]{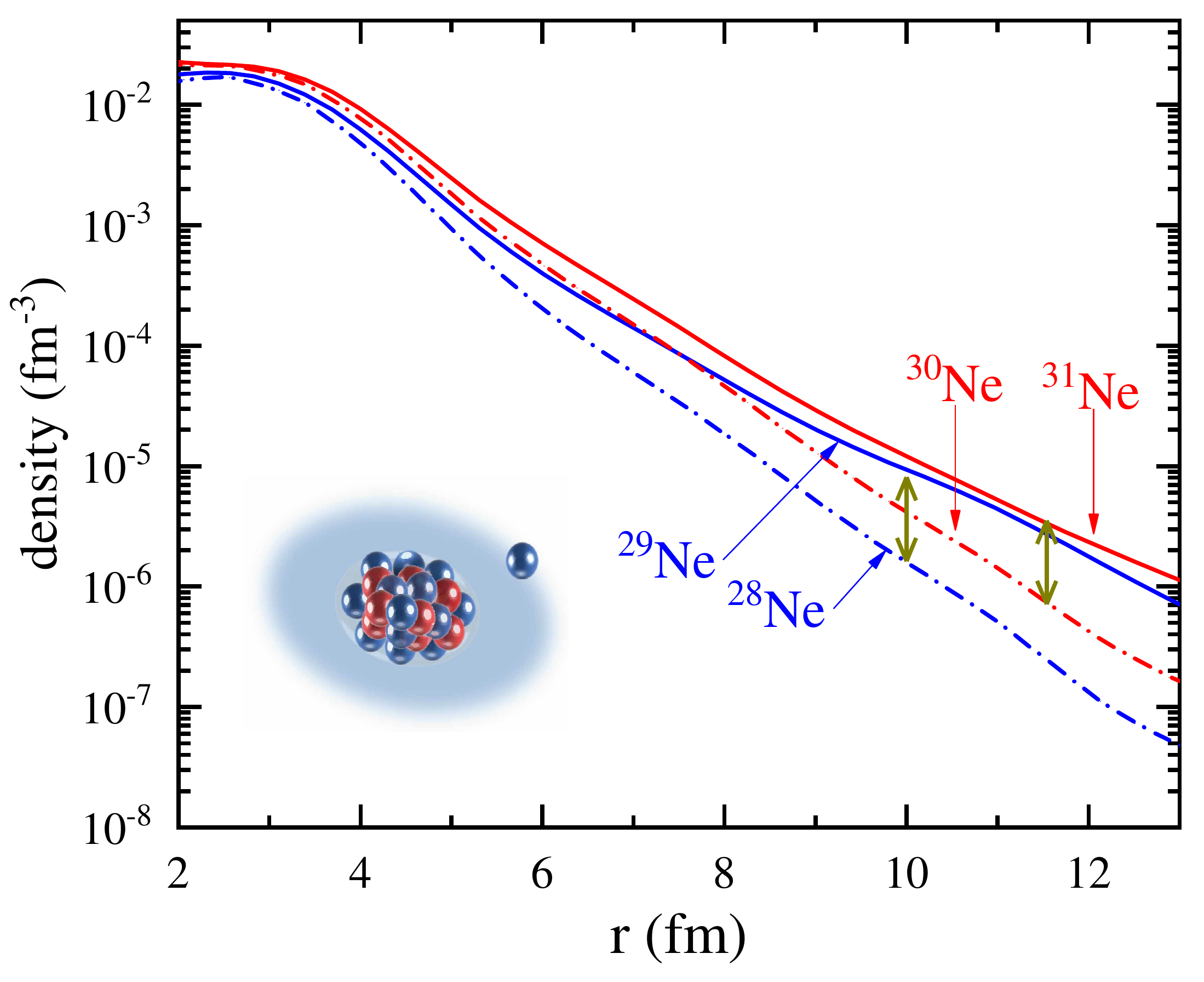}
\caption{The valence-nucleon densities of the $^{28-31}$Ne isotopes calculated with GSM using the optimized pionless EFT interaction.
  The one-body densities of $^{29,31}$Ne are depicted by solid lines, while the one-body densities of the inner \textit{cores} $^{28,30}$Ne are illustrated by dot-dashed lines (color online).}\label{density}
\end{figure}

In one-neutron halos, the neutron density distributions are extremely extended when compared to those generated by their associated inner \textit{cores} in the asymptotic region. 
To analyze the one-neutron halo structures in $^{29,31}$Ne isotopes, we calculated the one-body densities of the ground states of the $^{29,31}$Ne isotopes and those of the corresponding inner \textit{cores} $^{28,30}$Ne with GSM.
The obtained results  are shown in Fig.~\ref{density}.
We can clearly therein see that the one-body density of the ground state of $^{31}$Ne decreases slowly when compared with that of its inner \textit{core} $^{30}$Ne in the asymptotic region.
Our above results can be seen to be similar to the one-body densities of the one-neutron \textit{s}-wave halo $^{15}$C \cite{PhysRevC.69.034613} and one-neutron \textit{p}-wave halo $^{37}$Mg \cite{PhysRevC.90.061305}.
The calculated one-body density of the $^{31}$Ne is also consistent with experimental data and supports the one-neutron \textit{p}-wave halo character of the $^{31}$Ne nucleus \cite{PhysRevLett.112.142501}.
A similar situation occurs in $^{29}$Ne, where the one-body neutron density also decreases slowly in the asymptotic region when compared to that of its inner \textit{core} $^{28}$Ne.
Therefore, the similar structure of the one-body densities of $^{29}$Ne and $^{31}$Ne suggests that $^{29}$Ne is a good candidate for one-neutron halo in the medium-mass region.

\begin{figure}[!htb]
\includegraphics[width=1.0\columnwidth]{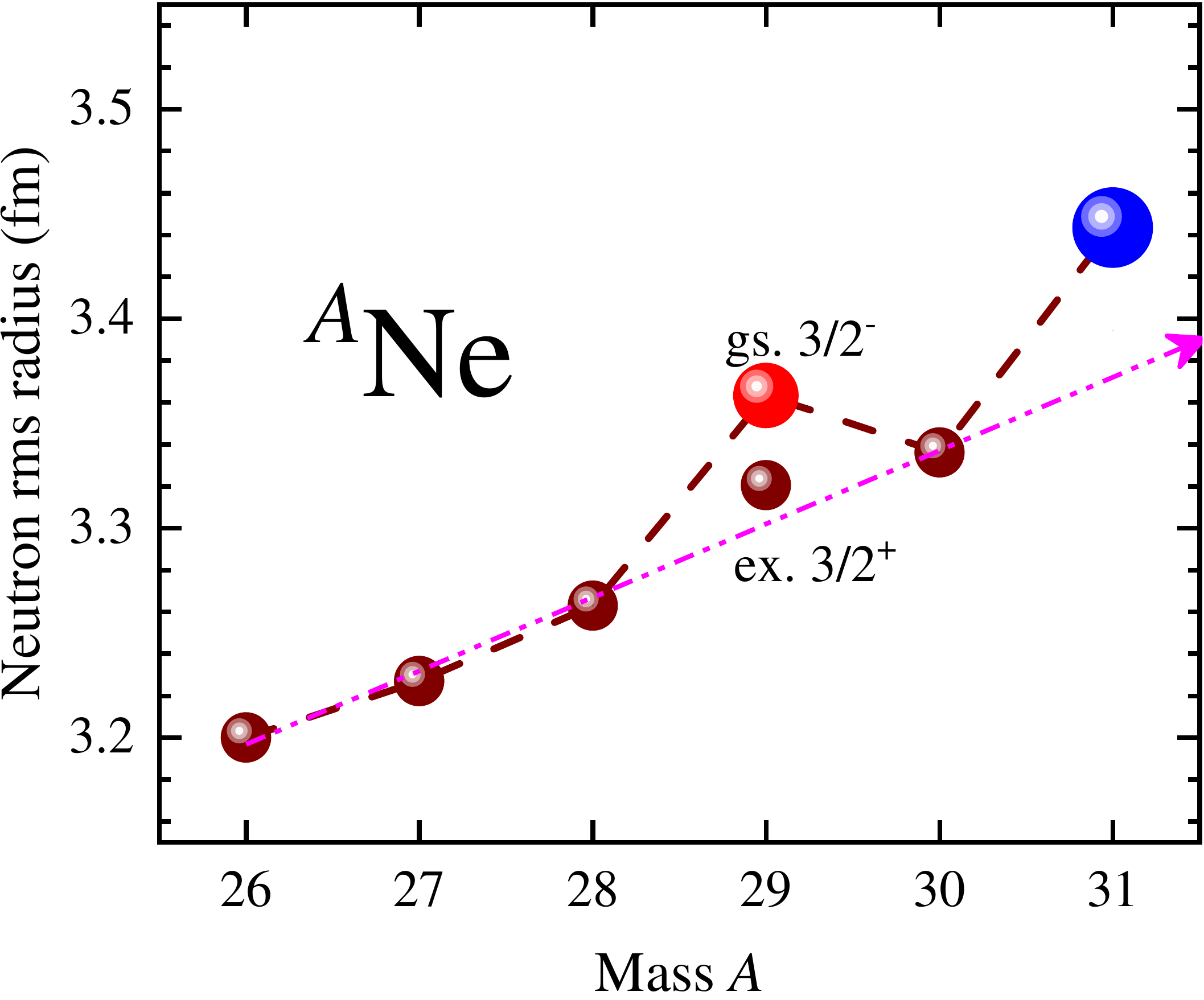}
\caption{ The calculated neutron rms radius of $^{26-31}$Ne with GSM. The dotted line ending with an arrow indicates the normal increasing trend of neutron rms radius as a function of neutron number (color online).}\label{radius}
\end{figure}

To further analyze the halo structures of $^{29,31}$Ne, and to provide deeper comparisons between the well-known one-neutron halo nucleus $^{31}$Ne with the halo candidate $^{29}$Ne,
we calculated the neutron rms radii of the ground states of $^{26-31}$Ne isotopes with GSM.
The results are presented in Fig.~\ref{radius}.
We can see that the calculated neutron rms radii of $^{29}$Ne and $^{31}$Ne do not follow the line formed by the neutron rms radii of other neon isotopes, $^{29,31}$Ne rms radii being situated 0.05 to 0.1 fm above this line.
The one-neutron halo of $^{31}$Ne is thus clearly shown by comparing its neutron rms radius to that of its inner \textit{core} $^{30}$Ne. Our results are in accordance with those of Ref.\cite{PhysRevLett.112.142501}.
A similar situation also occurs in the one-neutron halo $^{19}$C, as the experimental rms radius of $^{19}$C is larger than that of its inner \textit{core} $^{18}$C \cite{PhysRevLett.117.102501}.
To assess the effect of nuclear structure on neon rms radii, the neutron rms radius of the first $3/2^+$ excited state of $^{29}$Ne has been calculated for comparison.
Indeed, the latter is mainly dominated by rather localized shell model configurations, where $d$-waves are mostly occupied \cite{sym13112167}. 
GSM calculations provide a neutron rms radius of the $^{29}$Ne ground state ($3/2^-$) which is larger than that of its first $3/2^+$ excited state.
The large radius difference between the $3/2^-$ ground state and $3/2^+$ excited state of $^{29}$Ne thus provides an additional argument in favor of the presence of a one-neutron $p$-wave halo  in the $^{29}$Ne ground state.

GSM results for neutron rms radii of neon isotopes, increasing from $^{26}$Ne to $^{30}$Ne, also suggest the disappearance of the shell closure \textit{N} = 20 in the neon chain.
GSM results follow the conclusions derived from mass and spectra data, as the trend of neutron rms radii of neon isotopes is akin to that in magnesium isotopes of the island of inversion \cite{PhysRevLett.108.042504}.
Furthermore, the obtained neutron rms radii of $^{29,31}$Ne are also consistent with experimental interaction cross sections, as they give significantly greater values in $^{29,31}$Ne than in neighboring nuclei \cite{TAKECHI2012357}. However, our calculation of the neutron rms radius of $^{27}$Ne is not consistent with the slight enhancement of interaction cross section seen in $^{27}$Ne, which may be due to the lack of configuration-mixing in the present calculations, using a $^{24}$O core.

In one-neutron halos, spectroscopic factors of the valence neutron above the inner \textit{core} are typically large, with a value comparable to those associated to s.p.~states close to double-magic nuclei.
As continuum coupling plays an essential role in the calculations of spectroscopic factors in weakly bound and unbound nuclei, GSM is a proper model for their evaluation  \cite{PhysRevC.75.031301}.
In fact, despite their non-observable nature, the study of spectroscopic factors in halo nuclei is crucial to understand the exotic properties of nuclear systems close to drip-lines.
One-nucleon radial overlap functions can also provide important information about one-neutron halo nuclei, due to the fact that they resemble a valence nucleon wave function weakly coupled to an inner \textit{core} \cite{PhysRevC.85.064320}.

In the GSM framework, the one-nucleon radial overlap function \cite{PhysRevC.75.031301} is defined as :
\begin{equation}
u_{\ell j}(r) = \frac{1}{\sqrt{2J_A + 1}}\sum_{\mathcal{B}} \langle \widetilde{\Psi_{A}^{J_A}} || a_{\ell j}^{\dagger}(\mathcal{B}) || \Psi_{A-1}^{J_{A-1}} \rangle \langle r \ell j | u_{\mathcal{B}} \rangle
\end{equation}
where $a_{\ell }^{\dagger}(\mathcal{B})$ is a creation operator associated with the s.p.~basis state $|u_{\mathcal{B}}\rangle$.
The tilde symbol above the bra-vector reminds that complex conjugation is absent from matrix elements when using the Berggren basis, which can be rigorously used only in rigged Hilbert spaces \cite{Michel_GSM_book}.
Spectroscopic factors can be computed via the relation :
\begin{equation}
S = \int_0^{+\infty} u_{\ell j}^2 (r) dr. \label{DF}
\end{equation}
Spectroscopic factors are complex in GSM. The interpretation of complex observables is that their real part corresponds to the statistical average of all made measurements,
while their imaginary part corresponds to their statistical uncertainty (see Ref.\cite{Michel_GSM_book} and references therein). In our calculations, the imaginary part of spectroscopic factors is very small, so that they can be neglected in our present analysis.
As the GSM calculations of one-nucleon overlap functions and spectroscopic factors are done by summing over all discrete Gamow states and discretized scattering states along the $L_+$ contour \cite{0954-3899-36-1-013101},
they are independent of the used s.p.~basis \cite{PhysRevC.75.031301}. This is not the case in HO shell model calculations, performed in a limited model space where usually only one state of a given partial wave is present \cite{RevModPhys.77.427}.

\begin{figure}[!htb]
\includegraphics[width=1.0\columnwidth]{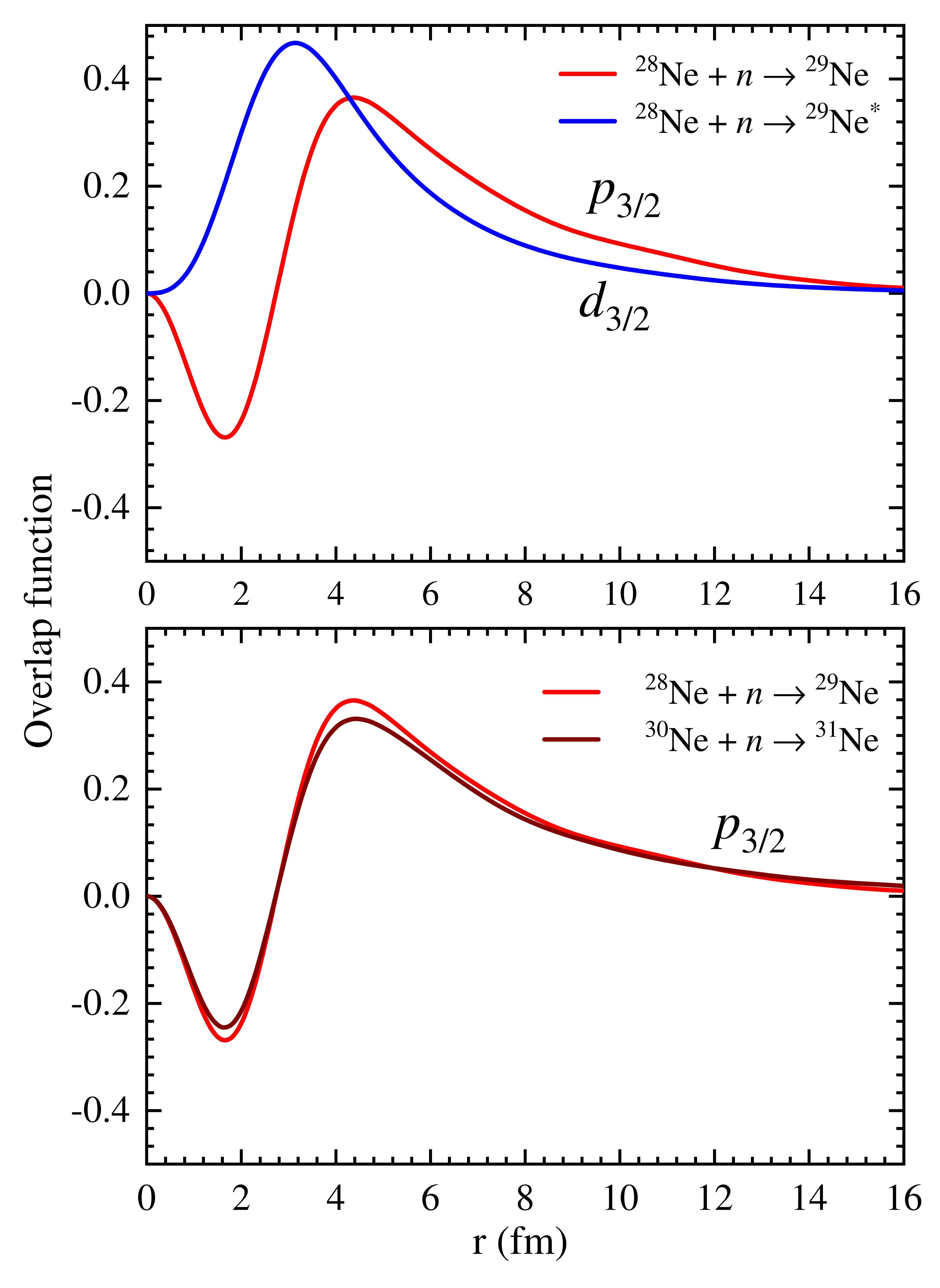}
\caption{The one-neutron overlap functions of the $^{29}$Ne and $^{31}$Ne calculated with GSM (color online).}\label{overlap}
\end{figure}
 
The one-neutron overlap functions of the ground states of $^{29,31}$Ne and the first ${3/2}^+$ excited state of $^{29}$Ne corresponding to the ground states of $^{28,30}$Ne are calculated. Results are shown in Fig.~\ref{overlap}. 
One-neutron halo structure can be clearly seen from the extended one-neutron overlap function in the asymptotic region.
As the ground states of $^{29}$Ne and $^{31}$Ne have many-body quantum numbers $J^\pi = 3/2^-$, they mainly consist of an inner \textit{core} coupled to a \textit{p}-wave valence neutron.
The first $3/2^+$ excited state of $^{29}$Ne, on the contrary, is dominated by the inner \textit{core} coupled to a \textit{d}-wave valence neutron.
The comparison of the one-neutron overlap function of $^{29}$Ne with the well-known halo $^{31}$Ne is essential for the predictions of the halo structure of $^{29}$Ne. 
We can see from Fig.~\ref{overlap} that the one-neutron overlap function associated to the neutron $p_{3/2}$ partial wave in the ground state of $^{29}$Ne is similar to that of $^{31}$Ne. 
Furthermore, the overlap function of the ground state of $^{29}$Ne shows a slower decrease than that of the $d_{3/2}$ one-neutron overlap function of the first $3/2^+$ excited state of $^{29}$Ne.
This situation arises as $p_{3/2}$ components are more extended in the asymptotic region than $d_{3/2}$ components because of their different centrifugal parts.
Hence, GSM one-neutron overlap functions also support the \textit{p}-wave one-neutron halo character of the ground state of $^{29}$Ne.

We calculated $p_{3/2}$ spectroscopic factors of $^{29,31}$Ne from the obtained  one-neutron overlap functions.
We obtained the values 0.540 and 0.392 for $^{29}$Ne and $^{31}$Ne, respectively. 
Both these results are close to experimental data, which are equal to 0.54(9) \cite{PhysRevC.93.014613} and $0.32^{+21}_{-17}$ \cite{PhysRevLett.112.142501}, respectively. 
The calculated neutron $p_{3/2}$ spectroscopic factor in $^{29}$Ne is larger than that of $^{31}$Ne, which is consistent with experiment data.
Nuclear deformation can be described via a configuration-mixing shell-model picture using a large valence space. Stronger $d_{3/2}$ and $pf$ cross-shell configuration mixing in the $^{30}$Ne ground state is obtained in our GSM calculation compared to the $^{28}$Ne ground state. Results suggest that deformation in $^{30}$Ne is larger than that in $^{28}$Ne. A similar situation is obtained in HO shell model calculations \cite{PhysRevC.93.014613}. Furthermore, a large deformation arising from a strong configuration mixing in the inner core of one-neutron halo nucleus typically provides a smaller spectroscopic factor value. The experimental and theoretical spectroscopic factors of $^{29,31}$Ne indicate that the deformation of the \textit{core} $^{28}$Ne in the one-neutron halo nucleus $^{29}$Ne is smaller than that of the \textit{core} $^{30}$Ne in the one-neutron halo nucleus $^{31}$Ne.

We checked that the obtained results do not qualitatively change with other parametrizations of the Hamiltonian. For this, one used the Hamiltonian of Ref.\cite{PhysRevC.101.031301}, where the one-body WS part was slightly refitted in order for the separation energies of $^{29,31}$Ne to be reproduced. The general trend observed in densities, rms radii and overlap functions (see Figs.~\ref{density},\ref{radius}, and \ref{overlap}) is indeed also present with the Hamiltonian of Ref.\cite{PhysRevC.101.031301}, these values being quantitatively different from those illustrated in this paper by about 10\% at most. Consequently, the one-neutron halo structure found in $^{29}$Ne is very likely to occur experimentally, as it is reproduced by two different Hamiltonians, where parameters are fitted only on energies and where densities, rms radii and overlap functions are predicted. Such a situation had already been encountered in Ref.\cite{PhysRevC.101.031301}, where part of the authors investigated the possibility of a two-neutron halo of $^{31}$F with GSM.

\textit{Summary.} --
Drip-line nuclei exhibit unique phenomena, such as halo structure. 
The multiconfigurational approach GSM, in which both many-body correlations and continuum coupling are included, has been used to study the one-neutron halo character of $^{29}$Ne and $^{31}$Ne. 
The one-neutron separation energies of $^{29,31}$Ne, which are essential in the formation of their halos, are well reproduced by using a Hamiltonian whose parameters are fitted from nearby nuclei.
One-body densities, neutron rms radii, and one-neutron overlap functions of neutron-rich neon isotopes have also been calculated. 
The well-known one-neutron halo $^{31}$Ne happens to be well described in our GSM calculations.
By comparing the calculated one-body density, neutron rms radius, and one-neutron overlap function of $^{29}$Ne, to those of the well known one-neutron halo $^{31}$Ne,
we could establish that $^{29}$Ne and $^{31}$Ne exhibit halo-dependent observables of similar value and shapes.
The neutron rms radius and one-neutron overlap functions of the first $3/2^+$ excited state in $^{29}$Ne have been calculated for comparison.
Their qualitative difference with those of the ground state of $^{29}$Ne also supports its one-neutron halo character.
As an additional test for that matter, spectroscopic factors involving the weakly bound valence neutron in $^{29,31}$Ne have been computed using the obtained one-neutron overlap function.
Results have been found to be consistent with experimental values.
Consequently, our calculations suggest that the ground state of $^{29}$Ne exhibits a one-neutron \textit{p}-wave halo, and, hence, that it is a good candidate for one-neutron halo nucleus in the medium-mass region.

\textit{Acknowledgments} -- 
 This work has been supported by the National Natural Science Foundation of China under Grant Nos. 11921006, 12175281 and 11975282; the Strategic Priority Research Program of Chinese Academy of Sciences under Grant No. XDB34000000; the Key Research Program of the Chinese Academy of Sciences under Grant No. XDPB15; the State Key Laboratory of Nuclear Physics and Technology, Peking University under Grant No. NPT2020KFY13.


\section*{References}

\bibliographystyle{elsarticle-num_noURL}
\bibliography{Neon_isotopes}





\end{document}